\documentclass[conference]{IEEEtran}

\IEEEoverridecommandlockouts
\usepackage{cite}
\usepackage{amsmath,amssymb,amsfonts}
\usepackage{algorithmic}
\usepackage{graphicx}
\usepackage{textcomp}
\usepackage{xcolor}
\usepackage{algorithmic}
\usepackage{algorithm}
\usepackage{array}
\usepackage[caption=false,font=normalsize,labelfont=sf,textfont=sf]{subfig}
\usepackage{textcomp}
\usepackage{stfloats}
\usepackage{url}
\usepackage{verbatim}
\usepackage{graphicx}
\usepackage{cite}
\usepackage{booktabs}
\usepackage{cite}
\usepackage{amsmath,amssymb,amsfonts}
\usepackage{algorithmic}
\usepackage{graphicx}
\usepackage{textcomp}
\usepackage{xcolor}
\usepackage{booktabs}
\usepackage{subfig}
\graphicspath{{Figures/}}

\usepackage[top=0.75in, bottom=1.1 in, left=0.673in, right=0.65in]{geometry}

\title{Resource Allocation for XR with Edge Offloading: A Reinforcement Learning Approach}

\author{\IEEEauthorblockN{ 
Alperen Duru\IEEEauthorrefmark{1},
Mohammad Mozaffari\IEEEauthorrefmark{2},
Ticao Zhang\IEEEauthorrefmark{2}, and
Mehrnaz Afshang\IEEEauthorrefmark{2} 
\vspace{0.2cm}}
\IEEEauthorblockA{\IEEEauthorrefmark{1}University of Texas at Austin, Austin, TX, USA\\ Email: aduru@utexas.edu}
\IEEEauthorblockA{\IEEEauthorrefmark{2}Ericsson Research, Silicon Valley, Santa Clara, CA, USA \\ Emails: \{mohammad.mozaffari, ticao.zhang, mehrnaz.afshang\}@ericsson.com} \vspace{-0.7cm}

}

\begin{document}
\maketitle

\begin{abstract}
Future immersive XR applications will require energy-efficient, high data rate, and low-latency wireless communications in uplink and downlink. One of the key considerations for supporting such XR applications is intelligent and adaptive resource allocation with edge offloading.  To address these demands, this paper proposes a reinforcement learning-based resource allocation framework that dynamically allocates uplink and downlink slots while making offloading decisions based on the XR headset's capabilities and network conditions. The paper presents a numerical analysis of the tradeoff between frame loss rate (FLR) and energy efficiency, identifying decision regions for partial offloading to optimize performance. Results show that for the used set of system parameters, partial offloading can extend the coverage area by 55\% and reduce energy consumption by up to 34\%, compared to always or never offloading. The results demonstrate that the headset's local computing capability plays a crucial role in offloading decisions. Higher computing abilities enable more efficient local processing, reduce the need for offloading, and enhance energy savings.
\end{abstract}


\section{Introduction}
With the growing number of extended reality (XR) applications, reliable wireless connectivity and efficient computing systems have become essential to meet the increasingly stringent performance requirements. Recognizing the importance of XR in this evolving landscape, it has been a key focus in the standardization of 5G Advanced, shaping the direction of next-generation wireless networks towards 6G \cite{3GPPTR26928, 3GPPTR38835, gapeyenko2023standardization, XR_beam}. Despite these advancements, current XR head-mounted displays still face significant challenges due to their limited battery capacity. To address these issues, edge computing has emerged as a promising solution to enhance energy efficiency by offloading resource-intensive tasks to servers, thereby reducing the processing burden on the device.

Future XR applications' unique demands, including uplink (UL) traffic and strict latency constraints, require specialized solutions. Delays in frame transmission and rendering beyond a set threshold can cause visual de-synchronization, leading to nausea or VR sickness for latencies on the order of a few tens of milliseconds \cite{3GPPTR26928}. The real-time nature of XR means that frames are lost if they exceed this latency threshold, with buffering not being a viable option. This makes FLR an important performance indicator. Balancing FLR and energy efficiency depends on effective communication resource allocation and edge offloading strategies. While offloading tasks to the edge can improve XR energy efficiency, it increases the communication load. Thus, optimizing resource allocation and offloading decisions play a crucial role in balancing computational demands with network load, achieving a tradeoff between FLR and energy consumption.


\subsection{Related Work}

Most prior studies on edge computing for XR applications primarily focus on offloading tasks without stringent latency requirements and predominantly target the downlink (DL) connection \cite{cui2021reinforcement, sun2020reducing}. These studies often overlook the importance of UL connections, which are becoming increasingly critical for emerging XR applications such as streaming environments in real-time \cite{gapeyenko2023standardization}. In contrast, this paper includes a numerical analysis of both UL and DL FLRs in relation to energy efficiency, while simultaneously considering slot allocation and partial offloading strategies for XR headsets. Additionally, while many approaches in the literature aim to minimize latency, they treat it as a parameter to be optimized rather than as a strict constraint that must be met on a frame-time interval basis \cite{ren2018latency, saleem2018performance}. In contrast, this study takes latency as a constraint to ensure that the frame transmission remains within acceptable bounds, directly addressing the requirements for real-time XR applications.

Several studies have explored task offloading to the cloud or edge but focus on offloading larger tasks that operate on timescales of several seconds, rather than the millisecond-level demands of XR frame rendering \cite{trinh2022deep, wang2024dynamic}. These approaches often involve intricate task handoff mechanisms, resulting in back-and-forth communication that does not align with the real-time requirements of XR systems \cite{yu2022quality}. In contrast, this paper proposes a dynamic partial offloading strategy that adapts to real-time conditions, meeting strict latency constraints while balancing energy efficiency effectively.

Moreover, while some recent works aim to optimize quality of experience (QoE) and frame quality in DL, they frequently neglect the impact of network bandwidth and headset computing capabilities on offloading decisions \cite{pan2024quality, nyamtiga2024adaptive}. Unlike these studies, this work explicitly analyzes how varying bandwidth and headset computing power affect the ideal offloading strategy, providing a detailed investigation into how these factors influence energy efficiency and offloading decisions. By integrating these considerations, this paper not only fills a gap in the literature but also sets a framework for designing more adaptive XR systems.


\subsection{Contributions}

The paper presents an analysis of time-division duplexing (TDD) resource allocation for partial edge offloading of future XR applications with UL/DL traffic, employing a reinforcement learning (RL) approach. The paper's main contributions are (1) an analysis of the effects of headset computing capabilities and allocated communications bandwidth on the optimal partial offloading ratios and (2) the combination of UL/DL TDD slot allocation combined with partial offloading in the frame-level latency requirements for future XR communications.  This paper employs a deep Q-network (DQN) to manage the allocation of uplink and downlink slots and make offloading decisions for an XR headset in an urban macro-cell (UMa) non-line-of-sight (NLOS) environment. The proposed RL method improves the coverage area by up to 55\% compared to always offloading, while improving the energy efficiency by up to 32\% compared to never offloading. The analysis shows that the headset computing capabilities and allocated bandwidth impact the FLR and energy efficiency tradeoff and optimal offloading decisions. 

\section{System Model}


A single XR user communicating with a single base station (BS) with the allocated UL/DL TDD slots is considered. The XR headset is equipped with cameras and uploads real-time video frames of its surroundings while sharing computation with the edge server co-located with the BS. Real-time communication with another user through streaming the environment by camera frames sets stringent latency thresholds on both UL and DL transmissions. The set of operations can be summarized as follows:
\begin{enumerate}
    \item XR headset starts uploading its camera frames.
    \item While the XR headset is uploading, part of the frame is rendered and encoded on the edge, utilizing the time spent in UL transmission.
    \item Once the UL frame is uploaded, the rendered frame is sent in the DL, if the rendering process is finished. Otherwise, the headset waits for the frame to arrive.
    \item The encoded DL frame is sent to the XR headset.
    \item The XR headset decodes the frame and continues executing from where the edge is left off.
    \item Frame is displayed and the process loops to Step 1.
\end{enumerate}

For ease of analysis in the system model, several assumptions are made. The transmission time for pose information in the uplink is considered negligible, with the data interpolated at the edge server based on past pose estimates. Frame sizes are modeled using a truncated Gaussian distribution as specified by 3GPP \cite{3gpp_tr_38_838}, and time correlations between frames are disregarded. Frames that exceed the latency threshold are immediately discarded and marked as lost. The communication is assumed to be a TDD between UL and DL transmissions. Next, we define different latency components in the considered XR model.

\subsection{Frame Latency}

Energy consumption on the XR headset arises from UL and DL communication, decoding, local execution, and idling time of the XR headset. Given the frame sizes and the power parameters, energy and latency for UL/DL transmission can be calculated, along with frame loss due to latency constraints. The latency and energy spent through the UL/DL time are connected through the power for the related power parameter. Therefore, defining the latency spent at each step is required for accurate energy computation and frame loss detection.

\subsubsection{Transmission Latency}
The latency for the frames consists of UL transmission, edge computing and encoding, DL transmission, local decoding of the pre-rendered frames at the XR headset, and finally, local computing. Each step needs to be defined clearly to compute the energy spent and detect frame losses due to latency constraints. For the frame transmission latency, the achievable rate should be defined for UL and DL. The Shannon rate for UL and DL transmission is computed with the channel gain at time $t$ for frame $f$, showing the individual slot and frame indices. The achievable data rates in UL and DL denoted by $R_t^{\mathrm{UL}}$ and $R_t^{\mathrm{DL}}$ can be found for the matched filter bound (MFB) of the channel as
\begin{equation}
\begin{split}
R_{f,t}^{\mathrm{UL}} &= B \log_2 \left( 1 + \frac{h_{f,t}^{\mathrm{MFB}} P^{\mathrm{UL}}}{B N_0} \right),\\
R_{f,t}^{\mathrm{DL}} &= B \log_2 \left( 1 + \frac{h_{f,t}^{\mathrm{MFB}} P^{\mathrm{BS}}}{B N_0} \right),
\end{split}
\label{R^ULDL}
\end{equation}
where $t$ is the slot number, $B$ is the bandwidth, $h_{f,t}^{\mathrm{MFB}}$ is the MFB of channel for the slot and frame indices, $N_0$ is the noise spectral density, and $P^{\mathrm{BS}}$ and $P^{\mathrm{UL}}$ are the BS and UL transmit powers, respectively. With the achievable data rates for UL and DL defined, the data size to be sent through DL needs to be defined as it will be affected by the offloading ratio decision. The data size in DL not only plays an important role in local and edge execution but also in the transmission time. For this reason, the size of data to be rendered on edge and locally for frame $f$ is defined as $D_f^{\mathrm{edge}}$ and $D_f^{\mathrm{loc}}$, defined by:
\begin{equation}
D_f^{\mathrm{edge}} = \alpha_f D_f^{\mathrm{DL}},
\label{Df^edge} 
\end{equation}
\begin{equation}
D_f^{\mathrm{loc}} = (1 - \alpha_f) D_f^{\mathrm{DL}},
\label{Df^loc}
\end{equation}
where $D_f^{\mathrm{DL}}$ stands for the size of the frame in DL. The transmission latencies are closely tied to the number of slots allocated to the UL and DL. The UL and DL rates in (\ref{R^ULDL}) determine the success of the frame transmission and contribute to the frame FLR. For slot time $\Delta t^{\mathrm{slot}}$ and the number of allocated time slots in UL and DL given by $N_f^{\mathrm{UL}}$ and $N_f^{\mathrm{DL}}$, the UL and DL transmission times are given by
\begin{equation}
L_f^{\mathrm{UL}} = N_f^{\mathrm{UL}} \Delta t^{\mathrm{slot}},
\label{Lf^UL} 
\end{equation}
\begin{equation}
L_f^{\mathrm{DL,comm}} = N_f^{\mathrm{DL}} \Delta t^{\mathrm{slot}}, \vspace{0.1cm}
\label{Lf^DLcomm}
\end{equation}
where $L_f^{DL,comm}$ is defined specifically for the communication part as the DL frame latency will also include the computation latency at the end.

\subsubsection{Computation Latency}
For the DL frame, there is also associated computation latency as pre-rendering and encoding on the edge, decoding and locally computing on the XR headset. The latency for edge computing and encoding can be computed by using \eqref{Df^edge} as $L_f^{\mathrm{edge}} = \frac{D_f^{\mathrm{edge}}}{F^{\mathrm{edge}}} + \frac{D_f^{\mathrm{edge}}}{F^{\mathrm{encode}}}$,
where $F^{\mathrm{edge}}$ and $F^{\mathrm{encode}}$ stand for the edge computing and encoding speeds, respectively.  The decoding in the XR headset also uses the offloaded data size in DL in \eqref{Df^edge} as $L_f^{\mathrm{decode}} = \frac{D_f^{\mathrm{edge}}}{F^{\mathrm{decode}}}$.
Once the pre-rendered part of the DL frame is received and decoded, XR headset needs to render the rest of the frame locally. The latency for local rendering can be computed by leveraging the remaining data size given in (\ref{Df^loc}) as $L_f^{\mathrm{loc}} = \frac{D_f^{\mathrm{loc}}}{F^{\mathrm{loc}}}$.
The total latency for transmission and computation is:
\begin{equation}
L_f^{\mathrm{DL}} = L_f^{\mathrm{UL}} + L_f^{\mathrm{DL,comm}} + L_f^{\mathrm{decode}} + L_f^{\mathrm{loc}}.
\label{Lf^DL}
\end{equation}

\subsection{Frame Energy}
The energy spent for UL and DL frame transmission and computation cycles depends on the individual delays of the steps. The communication framework parallelizes the UL transmission with the edge computing and encoding of the DL frame. 
If the edge computing and encoding take more time than the UL transmission, the XR headset waits in the idle state until the DL transmission starts with an upcoming DL slot to save energy. Hence, the energy spent during the camera frame uploading parallelized with edge computing and encoding for frame $f$ is given by
\begin{equation}
    E_f^{\mathrm{UL}} = L_f^{\mathrm{UL}} P^{\mathrm{UL}},
\end{equation}
\begin{equation}
E_f^{\mathrm{UL, edge}}=
\begin{cases} 
    E_f^{\mathrm{UL}}, & \hspace{-0.7cm}\text{if\,\,} L_f^{\mathrm{UL}} > L_f^{\mathrm{edge}}, \\ 
    E_f^{\mathrm{UL}} + \left\lceil \frac{L_f^{\mathrm{edge}} - L_f^{\mathrm{UL}}}{\Delta t^{\mathrm{slot}}} \right\rceil \Delta t P^{\mathrm{idle}}, & \text{otherwise}
\end{cases}
\label{Ef^ULedge}
\end{equation}
where $E_f^{\mathrm{UL}}$ is the energy spent in UL transmission, $P_f^{\mathrm{UL}}$ and $P^{idle}$ are the UL and idle powers of the XR headset, respectively. The ceiling function counts the number of affected DL slots due to extended latency in the edge computing and encoding process. Based on how many of the DL slots are affected, the energy in the DL communication is affected, which can be defined by
\begin{equation}
    E_f^{\mathrm{DL}} =
    \begin{cases} 
        N_f^{\mathrm{DL}} \Delta t P^{\mathrm{DL}}, & \hspace{-0.2cm}\text{if } L_f^{\mathrm{UL}} > L_f^{\mathrm{edge}} \\ 
        \left[N_f^{\mathrm{DL}} - \left\lceil \frac{L_f^{\mathrm{edge}} - L_f^{\mathrm{UL}}}{\Delta t} \right\rceil \right] \Delta t P^{\mathrm{DL}}, & \text{otherwise}
    \end{cases},
    \label{Ef^DL}
\end{equation}
where $P^{\mathrm{DL}}$ stands for the DL power of XR headset. The slots affected by prolonged edge computing and encoding are not used in DL and hence, they are not included in the DL energy. Once the frame is received in DL, it needs to be decoded locally to continue the rendering on XR, which is defined as
\begin{equation}
E_f^{\mathrm{dec}} = \frac{D_f^{\mathrm{edge}}}{F^{\mathrm{dec}}} P^{\mathrm{dec}},
\label{Ef^dec}
\end{equation}
where $F^{\mathrm{dec}}$ and $P^{\mathrm{dec}}$ are the decoding speed and energy of the XR headset, respectively. Finally, the local computing energy can be computed by the local computing latency and local execution power $P^{\mathrm{loc}}$ as $E_f^{\mathrm{loc}} = L_f^{\mathrm{loc}} P^{\mathrm{loc}}$.

By combining all the energy terms, the total spent energy during frame $f$ is computed as
\begin{equation}
E_f = E_f^{\mathrm{UL,edge}} + E_f^{\mathrm{DL,comm}} + E_f^{\mathrm{dec}} + E_f^{\mathrm{loc}}.
\end{equation}

\subsection{Frame Loss Indicator Definition}
FLR is a key metric for XR communications due to the real-time nature of frame generation. Frames are lost if they exceed the latency threshold, leading to a reduction in quality, even if various concealment techniques are employed. The FLR is derived from the frame loss indicator (FLI) for UL and DL frames averaged over a time interval. To determine if the allocated time slots are sufficient to send the UL and DL frames, the amounts of data in UL and DL are given by: \vspace{-0.1cm}
\begin{equation}
D_f^{\mathrm{UL,sent}} = \sum_{t=1}^{N_f^{\mathrm{UL}}} R_{f,t}^{\mathrm{UL}}, \vspace{-0.1cm}
\label{Df^ULsent}
\end{equation}
\begin{equation}
D_f^{\mathrm{DL,sent}} = \sum_{t=N_f^{\mathrm{UL}}+1}^{N_f^{\mathrm{UL}}+N_f^{\mathrm{DL}}} R_{f,t}^{\mathrm{DL}},
\label{Df^DLsent}
\end{equation}
where (\ref{Df^DLsent}) is adjusted depending on the DL slots that are affected by prolonged edge computing and encoding. For each frame $f$, the UL and DL FLIs are defined based on the frames' transmission success within the allocated time slots and if the total latency is below the latency threshold. UL frames are lost if the allocated slots are insufficient for transmission. DL frames can be lost due to insufficient slots or if the total latency, including UL/DL transmission and edge rendering, exceeds the latency threshold. The FLIs for UL and DL are then denoted as $\ell_f^{\mathrm{UL}}$ and $\ell_f^{\mathrm{DL}}$, respectively.


\subsection{Uplink Power Adjustments}

To optimize energy consumption, the XR user adjusts UL power according to channel conditions. Starting with maximum power for safe transmission, the user estimates the required power for the next slots, using a discrete set of power levels. The smallest power level that satisfies the transmission requirement is selected, as defined by: \vspace{-0.1cm}
\begin{equation}
P_f^{\mathrm{UL}} = \min_{0 < \beta \leq 1} \beta P^{\mathrm{UL,max}} \quad \text{s.t. } D_f^{\mathrm{UL,sent}} \geq D_f^{\mathrm{UL}}.
\end{equation}

\section{Problem Formulation and Reinforcement Learning Solution}

\subsection{Problem Formulation}

The primary objective for the XR user is to minimize the FLR and energy consumption simultaneously, achieved through slot allocation and partial edge offloading. The quality of experience QoE is directly tied to FLR since it reflects the user’s connection to latency. The QoE can be expressed as
\begin{equation}
    q_f = - \sigma (\ell_f^{\mathrm{UL}} + \ell_f^{\mathrm{DL}}),
\end{equation}
where $\sigma$ is a balance parameter that weighs the FLR against normalized energy. To incorporate energy into the objective function, the frame energy is normalized either by the maximum possible energy consumption during transmission or by the average energy spent in the system, ensuring a reasonable balance with FLR. The overall goal is to minimize FLR while reducing energy consumption per frame.

The key optimization parameters are the number of UL and DL slots, $N_f^{\mathrm{UL}}$ and $N_f^{\mathrm{DL}}$, and the offloading ratio $\alpha_f$, where $\alpha_f \in [0, 1]$. The objective function can then be formulated as:

\begin{equation}
\max_{N_f^{\mathrm{UL}}, N_f^{\mathrm{DL}}, \alpha_f} \left( - \sum_{x=f}^{f + N^{\mathrm{window}} - 1} \left[ \sigma \tilde{q}_x + (1 - \sigma) \tilde{E}_x \right] \right),
\label{ObjectiveFunc}
\end{equation}
where $\tilde{q}_x$ is the summation of FLIs for UL and DL, $\tilde{E}_x$ is the normalized energy for frame $x$, and $N^{\mathrm{window}}$ is the time window for operation or episode length.

The constraints for the optimization are:
\begin{equation}
    \begin{aligned}
    C_1: & \ \alpha_f \in [0, 1], \\
    C_2: & \ N_f^{\mathrm{UL}} + N_f^{\mathrm{DL}} \leq N_f^{\mathrm{total}}.
    \end{aligned}
\end{equation}
Here, $N_f^{\mathrm{total}}$ represents the total number of available slots, and $\sigma$ controls the tradeoff between FLR and energy consumption: $\sigma = 1$ focuses solely on minimizing FLR, while $\sigma = 0$ prioritizes energy efficiency.

This problem is challenging to solve without prior knowledge of channel conditions for upcoming frame intervals. Decisions on TDD allocation and offloading ratios must account for the variability and direction of channel changes, which remain uncertain before decision-making. When TDD and offloading actions are chosen at each frame interval, the state space expands exponentially, complicating the optimization of energy and FLR over time. As a result, the optimization tends to be conservative or hard to solve.

\subsection{Reinforcement Learning Approach}
The objective function is well suited for optimization using reinforcement learning (RL) due to the combinatorial problem's complexity. RL offers flexibility, enabling the adaptation of state and reward functions when introducing new parameters, unlike traditional methods that may require re-engineering the entire solution. A DQN is employed to estimate Q-values and serve as a baseline for numerical analysis, though alternative RL models can address the problem.

State: The state space is represented by a 3-dimensional vector $(D_f^{\mathrm{UL}}, D_f^{\mathrm{DL}}, h_f)$, where $D_f^{\mathrm{UL}}$ and $D_f^{\mathrm{DL}}$ are the UL and DL frame sizes, and $h_f$ is the channel gain at the beginning of frame $f$. This state information allows the agent to make informed decisions about slot allocation and offloading.

Action: The action space consists of the allocation decisions for UL and DL slots, $N_f^{\mathrm{UL}}$ and $N_f^{\mathrm{DL}}$, respectively, and the offloading ratio $\alpha$, where $\alpha \in [0, 1]$. These actions determine how the available resources are allocated between UL and DL and the portion of the frame that is processed locally versus offloaded to the edge server.

Reward: The reward function is designed to balance the FLR and energy efficiency. It is derived from \eqref{ObjectiveFunc} which is
\begin{equation}
    \text{Reward} = - \sigma \tilde{q}_x - (1 - \sigma) \tilde{E}_x,
    \label{RewardFunc}
\end{equation}
where $\tilde{q}_x$ is the FLR for UL and DL, and $\tilde{E}_x$ is the normalized energy for frame $x$. The parameter $\sigma$ determines the tradeoff between minimizing FLR and optimizing energy consumption. Care must be taken to ensure $\sigma$ is not too low, as this could lead the agent to avoid sending or receiving data altogether, which is undesirable. At the extreme case of $\sigma$ equal to 0, no resource is allocated to the user and the user simply waits in idle mode, not caring about the FLR.

The DQN employs a single hidden layer, normalizing input state variables based on data frame sizes and channel gain magnitudes. The output layer includes discretized offloading ratios and slot allocations, making workload splitting with the edge more practical by avoiding continuous ratios. The DQN predicts a reward based on UL and DL frame loss indicators and the energy expenditure of each action, training itself to maximize the reward defined in \eqref{RewardFunc}.

While the RL agent could adjust slot allocations and offload at every communication frame, such frequent adjustments can be impractical. The BS may not be configured to update the allocated time slots at every communication time frame. To address this, multiple DQN agents are trained in parallel, each optimizing over a distinct decision-update frequency to maximize different time windows in \eqref{ObjectiveFunc}. Depending on the selected window length, the appropriate agent makes the decisions, tailoring decision conservatism to channel, data frame size variability, and specific time window requirements.

This DQN-based RL approach takes in the channels as input to learn the channel variations. The channel changes either due to XR headset motion or small scale fading. The RL approach learns to make the optimal decisions to balance the FLR and energy efficiency. This allows the system to react to channel variations proactively and efficiently allocate resources and make offloading decisions in dynamic network environments.

\section{Results and Discussions}

The wireless channels were simulated for a single XR UE and BS using the QuaDRiGa channel simulator in a UMa NLOS scenario. The simulations were carried out using the parameters listed in Table \ref{tab:HeadsetWirelessParams}. The UL and DL frames were generated at each time step according to a truncated Gaussian random variable as specified in 3GPP, to realize the appropriate frame sizes. The frame sizes were adjusted to meet the expected data rates for the XR application, approximately 28 Mbps for DL and 8.5 Mbps for UL. These values were generated in MATLAB and then imported into the Python environment for DQN algorithm simulations.

\begin{table}[htbp]
   \scriptsize
    \centering
    \caption{Evaluation parameters}
    \begin{tabular}{|l|c|}
        \hline
        \textbf{Parameter} & \textbf{Value} \\
        \hline
        \multicolumn{2}{|c|}{\textbf{Headset Power and Capabilities}} \\
        \hline
        Local Exec. Power & 0.5 W \\
        Local Decode Power & 0.1 W \\
        Max UL Power & 0.2 W \\
        DL Power & 0.3 W \\
        Idle Power & 0.001 W \\
        Local Computation Speed & 200 Mbps \\
        Local Decode Speed & 3 Gbps \\
        Edge Computation. Speed & 600 Gbps \\
        Edge Encode. Speed & 3 Gbps \\
        \hline
        \multicolumn{2}{|c|}{\textbf{Wireless Comm. Parameters}} \\
        \hline
        Carrier Frequency & 7 GHz \\
        Subcarrier Spacing & 15 kHz \\
        Bandwidth & 20 MHz \\
        BS Antenna Height & 25 m \\
        UE Height & 1.8 m \\
        Tx Power & 23 dBm/MHz \\
        Scenario & 3GPP UMa NLOS \\
        Average Distance & 500 m \\
        Antenna Config. & (8, 8) \\
        UE Antenna Config. & (2, 2) \\
        Deployment & Outdoors \\
        UE Speed & 5 km/h \\
        UE Rotations & Pitch = $\pi/4 \sin(\pi t)$, Yaw = $\pi/4 \cos(\pi t)$ \\
        Antenna Model & 3GPP \\
        Noise Figure & 7 dB \\
        Thermal Noise & -174 dBm/Hz \\
        BS Antenna Down Tilt & 12° \\
        \hline
    \end{tabular}
    \label{tab:HeadsetWirelessParams}
\end{table}

In the RL simulations, the environment was modeled with the reward function as defined in \eqref{RewardFunc}. During reward calculations, the headset power values from Table \ref{tab:HeadsetWirelessParams} were incorporated to compute the FLR and energy. These FLR and energy values were then transformed into rewards for the RL agent, which acted as feedback to guide its decisions. This corresponds to the XR UE computing its energy consumption and determining whether a frame was lost or not, then informing the BS. After each decision loop, new actions from the RL agent were sent to the BS and the edge server.

\subsection{Effects of Device Capability on Offloading Decisions}

\begin{figure}[!t]
    \centering
    \includegraphics[width=0.42\textwidth]{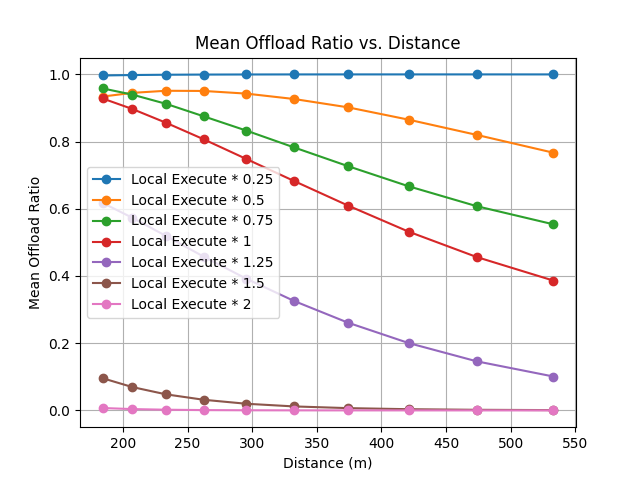}
    \caption{Mean offloading ratio as a function of distance for different local execution capabilities.}
    \label{fig:Offload_mean_vs_dist_runs_Finer}
\end{figure}

An important observation in this paper is how the XR device's local computing capability impacts the offloading decisions made by the RL agent. Fig. \ref{fig:Offload_mean_vs_dist_runs_Finer} illustrates how the mean offloading ratio changes with distance for different local execute capability scales. When the XR device has higher local processing power per spent energy, offloading becomes less frequent since local execution becomes more efficient. This suggests that the feedback from the XR headset's computing capability can play a key role in offloading decisions.

As the distance from the base station increases, there is a general trend of reduced edge offloading. Resources are allocated to uplink, and tasks are increasingly executed locally. This enables the RL agent to balance energy consumption and FLR more effectively, resulting in a better reward when the headset has higher computational capabilities. Fig. \ref{fig:Offload_mean_vs_dist_runs_Finer} suggests that as the local computing capabilities of the device change, the offloading decisions change significantly. 

\subsection{Coverage and Energy Efficiency Improvements}

The conventional baselines used for comparison with the proposed partial offloading method are the always offloading and never offloading scenarios. In all cases, the reward balancing is the same across methods and baselines only try to balance the allocated time slots to UL and DL.

\begin{figure}[!t]
    \centering
    \includegraphics[width=0.40\textwidth]{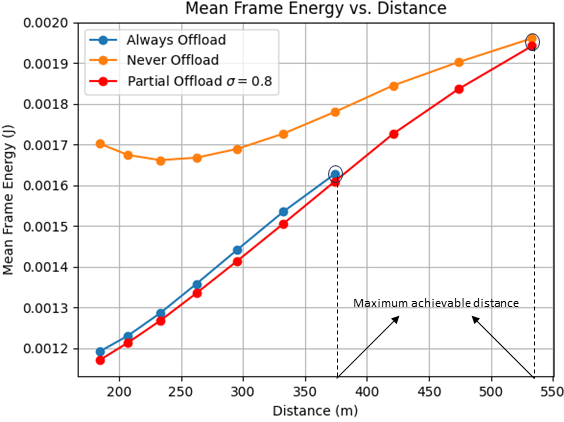}
    \caption{Mean energy consumption as a function of distance for different offloading methods. \vspace{-0.2cm}}
    \label{fig:Energy_mean_vs_dist_Finer3}
\end{figure}

\begin{figure}[!t]
    \centering
    \includegraphics[width=0.37\textwidth]{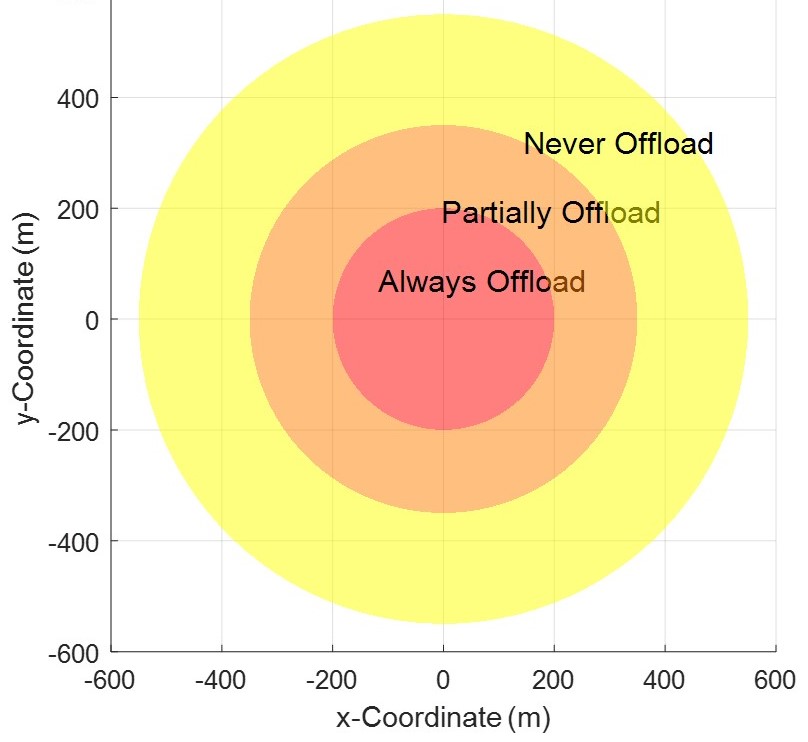}
    \caption{Example of offloading decision regions based on the distance.}  \vspace{-0.2cm}
    \label{fig:Offloading_region}
\end{figure}

Fig. \ref{fig:Energy_mean_vs_dist_Finer3} illustrates the mean frame energy as a function of distance for each algorithm plotted only up to the point where the FLR surpasses the specified threshold. Plots in Fig. 2 show only up to distances that can satisfy FLR limit of 10\%, where the \emph{Always Offload} case reaches up to 375 m while the other cases reach to 530 m. Compared to the always offloading scenario and for the given set of parameters, the partial offloading method achieves approximately 24\% greater coverage distance and around 55\% larger area coverage. In contrast, never-offloading and partial-offloading exhibit similar coverage capabilities by focusing on allocating more resources to UL to supply coverage, sacrificing energy. In terms of energy efficiency, partial offloading uses up to 32\% less energy than the never-offloading method while maintaining comparable coverage. This suggests that partial offloading balances the advantages of both always offloading and never offloading by extending coverage while simultaneously conserving energy. This makes partial offloading the most efficient option in terms of both coverage and energy consumption. \vspace{-0.02cm} Fig. \ref{fig:Offloading_region} shows an illustrative example of simplified offloading decision regions based on coverage distance. In practice, in a good coverage condition, always offloading can be a close to optimal decision, while in a poor coverage condition, the never offloading approach can be considered. For other cases, partial offloading provides the best performance.

\subsection{Effects of Bandwidth on Offloading Decisions}

\begin{figure}[!t]
    \centering
    \includegraphics[width=0.44\textwidth]{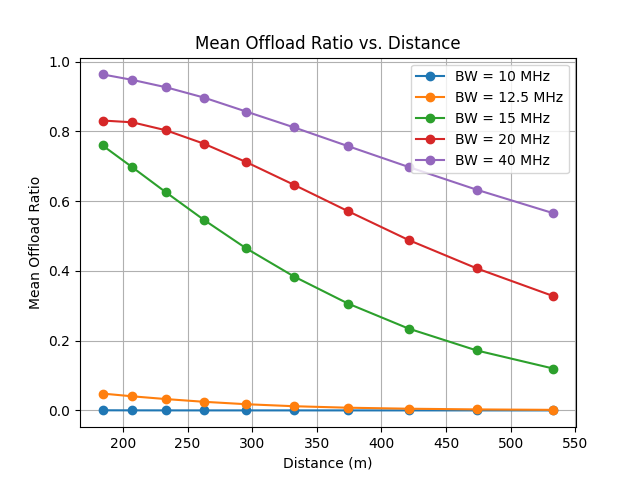}
    \caption{Mean offloading ratio as a function of distance for varying bandwidths. \vspace{-0.3cm}}
    \label{fig:Offload_mean_vs_dist_runs_bandwidth_dist_Finer}
\end{figure}

The amount of available bandwidth plays a significant role in offloading decisions. Fig. \ref{fig:Offload_mean_vs_dist_runs_bandwidth_dist_Finer} shows that the increased bandwidth enables the XR device to utilize edge server resources more efficiently, leading to a higher offloading ratio. This allows the system to offload more tasks to the edge, reducing energy consumption and maintaining a balance with the FLR. With greater bandwidth, the system gains more flexibility to offload tasks to the edge, leading to energy savings and better optimization of the FLR. This effect mirrors the impact of headset capability, where higher local processing power reduces the need for offloading, but with more bandwidth, offloading becomes more efficient.

\section{Conclusion}

This paper analyzed an envisioned XR application's communication requirements, focusing on partial offloading with UL/DL TDD allocation. Optimal strategies were identified based on distance, device capability, and bandwidth, revealing key factors for energy efficiency and FLR. Device capability strongly shapes offloading, as advanced XR headsets reduce reliance on edge computing. Numerical analysis highlighted the sensitivity of the offloading decisions to the device computing capabilities. Increasing bandwidth improves offloading efficiency up to a limit, offering guidance for resource allocation. Partial offloading outperformed always and never offloading, extending coverage distance by 24\%, area coverage by 55\%, and energy efficiency by up to 32\%. These results demonstrate the potential of partial offloading to balance energy use and FLR across scenarios. Future work may extend this framework to joint multi-user allocation, enabling adaptive UL/DL strategies for diverse user needs.


\def\baselinestretch{0.99}

\bibliographystyle{IEEEtran}
\bibliography{ref.bib}
\end{document}